\documentclass[pre,two column,notitlepage]{revtex4-1}
\usepackage[utf8]{inputenc}

\date{\today}
\usepackage{natbib}
\usepackage{graphicx}
\usepackage{amsmath}
\usepackage{amssymb}
\usepackage{comment}
\usepackage{csquotes}
\usepackage{xcolor}
\usepackage[compat=1.0.0]{tikz-feynman}
\usepackage[caption=false]{subfig}
\usepackage{relsize}

\newcommand{\phitilde}{\widetilde{\phi}}
\newcommand{\deltabar}{\bar{\delta}}

\begin{document}
\title{Generating and evaluating 1PI Feynman diagrams for an MSR field theory}
\author{Rajiv G. Pereira}
\affiliation{School of Physics, IISER Thiruvananthapuram, Vithura, Kerala 695551, India}

\begin{abstract}
We discuss certain computational methods and analytical techniques that can be used to automatize the various steps involved in  the renormalization group  analysis of an MSR field theory. The methods rely mainly on the well known packages FeynArts and SecDec. The former is used with minor modifications for generating the Feynman diagrams and obtaining the corresponding expressions, and the latter for dimensional regularization and epsilon expansion of the Feynman integrals. We first discuss how suitable classes and generic model files required by FeynArts are created, present the minor modifications made in the internal file \texttt{Analytic.m}, and then show how the diagrams and the expressions are generated. We then discuss the procedure followed in further simplifying the integrals obtained from the Feynman diagrams to render them in the form of standard scalar Feynman integrals so that the package SecDec can be used to dimensional regularize and epsilon expand them. We discuss these methods as it is applied to a particular theory, namely, the NSAPS three-vector model. However, they can be easily generalized to be applied to other similar MSR field theories. 
\end{abstract}

\maketitle

\section{Introduction}
Field theory techniques that involve Feynman diagrams and regularization of the integrals obtained therefrom are essential to various calculations in several areas of physics. It is well known that they are widely used in particle physics for scattering cross section and amplitude calculations~\citep{polchinski_1998, Peskin_2018}. In statistical mechanics, the renormalization group (RG) theory, which entails these techniques, is one of the most commonly used and robust theoretical methods of studying the critical scaling behavior of systems with large degrees of freedom~\citep{tauber_2014, Halperin_1977}.

The statistical properties of a system that follows Langevin dynamics can be studied by recasting the theory in Martin Siggia Rose (MSR) formalism~\citep{Martin_1973}. In this formalism, the statistical weight associated to an order parameter configuration can be written as the exponential of an action, which is referred to as the MSR action. Within this frame work,  we can use the standard field theory and RG techniques to extract relevant physical information about the system~\citep{tauber_2014, Zinn_2002}. These techniques are particularly useful in analyzing the critical scaling properties of various models and have been extensively used for the same over the years~\citep{tauber_2014, Janssen_1986, Janssen_1981, Bassler_1994, Schmittmann_1995}. However, the RG analysis can be a formidable task when the calculations involve numerous multi-loop Feynman diagrams. Computational methods are more desirable in such scenarios.   

 Several computational packages are available that can perform various tasks involved in the RG analysis of a typical quantum field theory. Packages like FeynArts~\citep{feynarts2001} and QGRAPH~\citep{qgraph1993} can be used to draw Feynman diagrams and obtain the respective expressions, while those like SecDec~\citep{secdec2015} and FIESTA~\citep{fiesta2009} can dimensional regularize and epsilon expand scalar Feynman integrals.   
 There are also several other programs like FeynCalc~\citep{feyncalc2016} and FARE~\citep{fare2016} that can perform various related tasks such as tensor reduction of Feynman integrals and Dirac algebra manipulations. 
 
 However, generating Feynman diagrams and obtaining the corresponding expression for an MSR field theory and regularizing the integrals therein computationally are not straightforward.  Though packages like FeynArts can generate Feynman diagrams for a quantum field theory they cannot be directly applied to an MSR field theory since the two point correlations are not Lorentz invariant and there are auxiliary fields which are different from the order parameter fields.  Even if the diagrams and the corresponding expressions are obtained, further simplifications are needed to convert the integrals to the form of standard Feynman integrals so that packages like SecDec can be used to dimensional regularize and epsilon expand them.   

In this paper, we discuss how various steps involved in the RG analysis of an MSR field theory can be automatized using a computer. For this we use the software Mathematica and the packages FeynArts and SecDec. These computational techniques were applied in a previous work to perform a two-loop RG analysis of the so called NSAPS three-vector model~\citep{pereira_2020}, which can be cast as an MSR field theory. Here, we discuss the computational techniques as it is applied to this model.  

We first describe how FeynArts can be used with minor modifications to generate the required 1PI Feynman diagrams and the corresponding expressions. We then discus the procedure for simplifying these expressions to render the integrals therein in the form of standard scalar Feynman integrals so that SecDec can be used to dimensional regularize and epsilon expand them. The procedure is such that it can be easily automated using Mathematica. Note that though the above procedures are discussed for the case of the NSAPS three-vector model, they can be easily generalized to be applied to other similar MSR field theories.     

 This article is organized as follows. In Sec.~\ref{sec:model}, we briefly describe the NSAPS three-vector model. Section~\ref{sec:draw} is dedicated to the computational methods used to generate the Feynman diagrams and the corresponding expressions, and Sec.~\ref{sec:dimreg} to the procedure to automatize dimensional regularization of the UV divergent integrals therein.  
\section{The NSAPS three-vector model}\label{sec:model}
We first recapitulate the field theoretic aspects of the NSAPS three-vector model that are essential to the forthcoming discussions. More details about this theory can be found in Refs.~\citep{pereira_2020, Dutta_2011}. The MSR action~\citep{Martin_1973} for the NSAPS three-vector model is written as 
\begin{align}\label{nsapsMSRreal}
S&=\sum_{a=1}^{3} \int_{x}\left[\widetilde{\phi}_{a}\left(\partial_{t}-D\left(\nabla_{\perp}^{2}+\rho \partial_{\|}^{2}-r\right)\right) \phi_{a}+ \frac{u_{0}}{3 !} \widetilde{\phi}_{a} \phi_{a}^{3} \right.
\nonumber
\\
&\left.
+\frac{u_{1}}{2 !} \widetilde{\phi}_{a} \phi_{a} \left( \phi_{a+1}^{2} + \phi_{a+2}^2 \right)
+e_{p} \phi_{a+1} \phi_{a+2} \partial_{\|} \widetilde{\phi}_{a}
-T \widetilde{\phi}_{a}^{2}\right] ,
\end{align}
where $x$ denotes the time and the space coordinates $\{t, \boldsymbol{x} \}$ and $\int_x \equiv \int dt d^d \boldsymbol{x}$. The subscripts $\perp$ and $\|$ denote the transverse and the longitudinal directions, respectively. The three component order parameter and auxiliary fields are denoted by $\phi_a$ and $\phitilde_a$, respectively, where $a \in \{1,2,3\}$. The parameter $T$ is the noise strength and $u_i$ and $e_p$ are coupling constants, the terms corresponding to which are treated perturbatively.   
The unperturbed action is written in Fourier space as
\begin{equation}
S_0=\sum_a \int_{q} \widetilde{\phi}_a(-q) \left(- i q_0 + M(\boldsymbol{q}) \right) \phi_a(q),
\end{equation}

\begin{figure}
\includegraphics[scale=0.23]{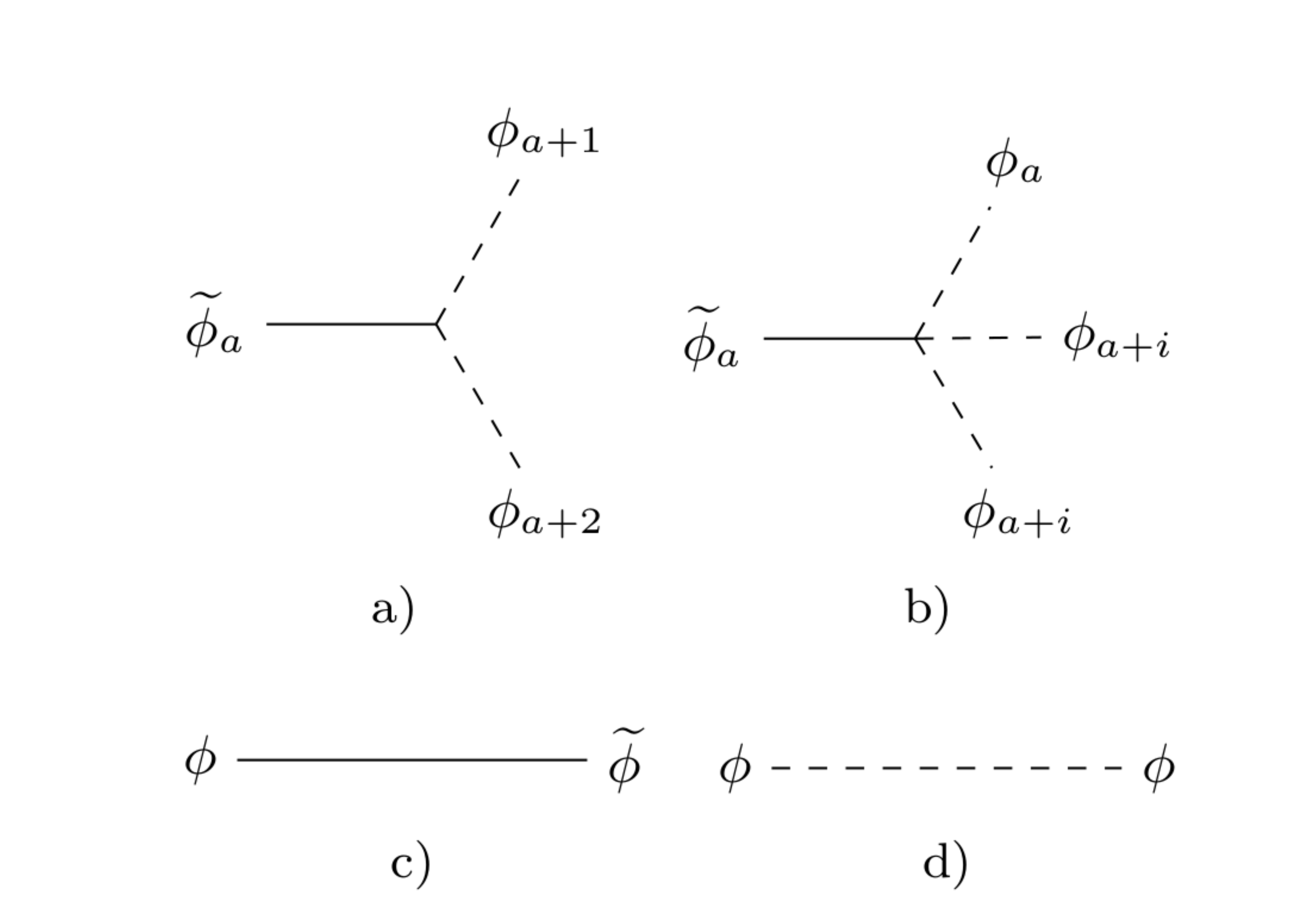}
\caption{The building blocks of Feynman diagrams for the NSAPS three-vector model. Figures~a) and b) show the diagramatic representation of the $e_p$ and $u_i$ interactions, respectively, given in Eq.~(\ref{nsapsInt}), and Figs.~c) and d) represent the propagators corresponding to $\langle \phi_a \phitilde_a \rangle_0$ and $\langle \phi_a \phi_a \rangle_0$, respectively.     }
\label{fig:buildingblocks}
\end{figure}
where $M(\boldsymbol{q})= D ({\boldsymbol{q}_{\perp}}^2 + \rho q_{\|}^2 + r)$ and $\int_{q} \equiv {\frac{1}{(2 \pi)^{d+1}}} \int dq_{0} d \boldsymbol{q}$. The Fourier transform of a function $f(x)$ is defined by the relation $f(x)=\int_q f(q) e^{-i q.x}$, where $q.x = q_0 x_0 -\boldsymbol{q} .\boldsymbol{x}$. The subscript $0$ denote the temporal direction. 

The two non-vanishing unperturbed two-point correlations are
\begin{align}\label{Gaussian_2pt}
& \langle \phi_a (q_1) \widetilde{\phi}_b (q_2) \rangle_0 = \frac{\delta_{ab} \bar{\delta}(q_1\!+\!q_2) }{-i q_{10} + M(\boldsymbol{q})}=\delta_{ab} \bar{\delta}(q_1\!+\!q_2) G_0(q_1),
\nonumber
\\
& \langle \phi_a (q_1) \phi_b (q_2) \rangle_0 = 2T\frac{\delta_{ab} \bar{\delta}(q_1\! +\!q_2) }{ q_{10}^2 + M(\boldsymbol{q})^2}\!=\!\delta_{ab} \bar{\delta}(q_1\!+\!q_2) C_0(q_1),
\end{align}
where $ \bar{\delta} (q) \equiv (2 \pi)^{d+1} {\delta}(q)$. 
The interaction part of the action is written in Fourier space as
\begin{widetext}
\begin{align}\label{nsapsInt}
  S_I=& \sum_{i=0}^2 \frac{u_i}{6} (3-2 \delta_{i,0} )\sum_{a=1}^3  \int_{q_1,...,q_4} \! \! \! \!\! \! \! \!\! \! \! \! \phitilde_a(q_1) \phi_a (q_2) \phi_{a+i} (q_3) \phi_{a+i} (q_4) \,  \deltabar(q_1 + q_2 +q_3+q_4) \nonumber \\
&+i e_p\sum_{a=1}^3  \int_{q_1,...,q_3} \! \! \! \!\! \! \! \!\! \! \! \! q_{1\|} \phitilde_a (q_1) \phi_{a+1}(q_2) \phi_{a+2} (q_3) \deltabar (q_1+q_2+q_3)
    \end{align}
\end{widetext}

The diagramatic representations of the interactions and the two point Gaussian correlations, which comprise the building blocks of the Feynman diagrams, are shown in Fig.~\ref{fig:buildingblocks}. Throughout this paper we adopt the convention that an external $\phi$ field hits from the left and an external $\phitilde$ field hits from the right. This makes the arrows which are usually drawn with the propagators redundant and are therefore discarded. 

The effective action is written as
\begin{equation}
\Gamma[\psi, \widetilde{\psi}]=-\ln\mathcal{Z}[J,\widetilde{J}] + \sum_a \int_x  J_a(x) \psi_a(x) + \widetilde{J}_a(x) \widetilde{\psi}_a(x),
\end{equation}
where
$
\psi(x)=\frac{\delta \ln \mathcal{Z}}{\delta J(x)}, \text{ }\textbf{   } \widetilde{\psi}(x)=\frac{\delta \ln \mathcal{Z}}{\delta \widetilde{J}(x)},
$
and the generating functional for correlation functions 
$
\mathcal{Z}[J,\widetilde{J}]=\left\langle \exp{\sum_a \int_x  \phi_a(x) J_a(x) + \widetilde{\phi}_a(x) \widetilde{J}_a(x)} \right\rangle.
$
The 1PI diagrams are obtained by taking the functional derivaties of $\Gamma$,

\begin{align}
\Gamma_{\widetilde{n},n}^{\widetilde{a}_1...\widetilde{a}_{\widetilde{n}}a_1 ... a_n}(\widetilde{x}_1,..& \widetilde{x}_{\widetilde{n}}; x_1,..x_n) =
\nonumber
\\
& \prod_{i=1}^{\widetilde{n}}\frac{\delta}{\delta \widetilde{\psi}_{\widetilde{a}_i} (\widetilde{x}_i)}
\left.\prod_{j=1}^{n} \frac{\delta}{\delta \psi_{a_j}(x_j)} \Gamma[\widetilde{\psi}, \psi]\right|_{\tilde{\psi}=\psi=0}.
\end{align}

The procedure of obtaining and evaluating the relevant 1PI Feynman diagrams computationally can be split into the following three steps. 
\begin{enumerate}
    \item Draw the relevant 1PI Feynman diagrams and obtain the corresponding integrals using the package FeynArts. 
    \item Write the integrals in the expressions obtained in step~1 in the form of standard scalar Feynman integrals.  
    \item Perform numerical dimensional regularization on these integrals using SecDec. 
\end{enumerate}
We now proceed to describe each of the above steps in detail.  

\section{Drawing Feynman diagrams and obtaining the expressions}\label{sec:draw}
In this section, we discuss how FeynArts is used to obtain the desired 1PI Feynman diagrams and the corresponding expressions for the NSAPS three-vector model. In particular, we first discuss how to create the model files, where the information about the model is stored, and the minor modifications in the internal file \texttt{Analytic.m} that are necessary to suit the package for our purpose. We then show how various functions in FeynArts are used to generate the diagrams and the expressions.    
Note that we do not repeat the instructions already given in the user manual of FeynArts~\citep{hahn} or descibe the techniques and algorithms thereof. For more details on these, see Refs.~\citep{FeynArtspaper, FeynArtspaper2, eck1995feynarts}. We discuss only the additional specifics and procedures
that are required to implement this calculation.

For convenience, we list below the important terminologies that will be used later in the discussion.  
\begin{itemize}
    \item \textit{Node} refers to a point where two or more lines meet. An \textit{n-node} is a point where n lines meet. 
    \item \textit{Topology} is a diagram that consists of lines and nodes. Unlike a Feynman diagram, a topology does not represent a mathematical expression. 
    \item \textit{Coupling} refers to any of the interaction terms in the action, in the case of the NSAPS three-vector model, any of the interaction terms in Eq.~(\ref{nsapsInt}). 
    \item \textit{Vertex} refers to the diagramatic representation of a coupling. For example, the vertices corresponding to the couplings given in Eq.~(\ref{nsapsInt}) are as shown in Fig.~\ref{fig:buildingblocks}. The term \textit{n-point} vertex refers to a vertex with \textit{n} lines meeting at a point. 
    \item \textit{Propagator} refers to any of the non-vanishing two-point Gaussian correlations without the Dirac and the Kronecker deltas. In particular, \textit{scalar} and 
    \textit{mixed} propagators refer to the ones corresponding to  $\langle \phi_i \phi_i \rangle_0$ and  $\langle \phi_i \phitilde_i \rangle_0$, respectively. 
\end{itemize}

\subsection{Creating model files}
The required information about the model is stored in two model files, the 
\textit{classes} and the \textit{generic} files. The classes file contains information about all the fields, the \textit{mixing} fields, and the couplings, and the generic file contains the information about the propagators. For further details and examples refer FeynArts manual~\citep{hahn}.   

We define each component of $\phi$ and $\widetilde{\phi}$ as a separate scalar field in the classes file. For instance, the fields $\phi_1$, $\phi_2$, and $\phi_3$ are defined as 
\begin{equation}\label{phiS}
\texttt{S[1], S[2],} \; \text{and} \; \texttt{S[3]},   
\end{equation}
respectively. The letter \texttt{S} stands for scalar, and the numbers in the bar brackets distinguish the different components of $\phi$. Likewise, the auxiliary fields $\phitilde_1$, $\phitilde_2$, and $\phitilde_3$ are defined as 
\begin{equation}\label{phitildeS}
 \texttt{S[11], S[12],} \; \text{and} \; \texttt{S[13]},     
\end{equation}
respectively. Each field is defined along with a set of attributes. While some of the attributes are optional the following three are mandatory and has to be set as indicated. 
\begin{itemize}
    \item \texttt{SelfCongugate} is set as \texttt{True} for both $\phi$ and $\phitilde$-fields. 
    \item \texttt{Mass} is set as \texttt{1} and \texttt{0} for the $\phi$ and the $\phitilde$ fields, respectively. Note that here we  \textit{do not} use the attribute \texttt{Mass} to indicate the mass of the particle. Instead, this option is used to identify which propagator is to be used when the \texttt{InsertFields} function is executed later.   
    \item \texttt{Indices} is set as \texttt{$\{ \}$} for both the $\phi$ and the $\phitilde$-fields. 
\end{itemize}
For example, the $\phi_1$ field is defined in the classes model file as follows. 
\begin{align}
    &\texttt{S[1] == \{} \nonumber \\
    &\texttt{SelfConjugate -> True,} \nonumber \\
    &\texttt{Mass -> 1,} \nonumber \\
    &\texttt{Indices -> \{\}} \nonumber \\
    &\texttt{\}}. 
\end{align}
There are a number of other auxiliary attributes that can be set according to the desired style of output Feynman diagrams. 

Apart from the $\phi$ and the $\phitilde$ fields, we also have to define the so called \textit{mixing} fields. A mixing field is defined for each of the non-vanishing Gaussian correlation involving two different fields. It is evident from Eq.~(\ref{Gaussian_2pt}) that there are three such correlations for the NSAPS three-vector model. However, separate mixing fields have to be defined for $\langle \phi_i \phitilde_i \rangle$ and $\langle \phitilde_i \phi_i \rangle$. Therefore we define six mixing fields in total, one corresponding to each of the following: $\langle \phi_1 \phitilde_1 \rangle$, $\langle \phi_2 \phitilde_2 \rangle$, $\langle \phi_3 \phitilde_3 \rangle$,  $\langle \phitilde_1 \phi_1 \rangle$, $\langle \phitilde_2 \phi_2 \rangle$, and $\langle \phitilde_3 \phi_3 \rangle$. As in the case of the $\phi$ and $\phitilde$-fields, mixing fields are defined with a number of attributes. While four of them are mandatory, the rest are optional. The mandatory attributes are set as described below.
\begin{itemize}
    \item \texttt{SeflCongugate} is set as \texttt{True}.
    \item \texttt{Mass} is set as \texttt{2}.
    \item \texttt{Indices} is set as $\{ \}$
    \item \texttt{MixingPartners} is set as \texttt{$\{$A,B$\}$}, where \texttt{A} and \texttt{B} are the fields involved in the non-vanishing Gaussian correlation the mixing field represents. For instance, the mixing field corresponding the correlation $\langle \phi_1 \phitilde_1 \rangle_0$ should have \texttt{MixingPartners} set as \texttt{\{S[1], S[11]\}}. Remember that \texttt{S[1]} and \texttt{S[11]} stands for $\phi_1$ and $\phitilde_1$, respectively, as already defined in Eq.~(\ref{phiS}) and (\ref{phitildeS}). 
\end{itemize}
The mixing field corresponding to $\langle \phi_1  \phitilde_1 \rangle_0$ is written in the classes file as
\begin{align}
    &\texttt{Mix[S,S][11] == \{}\nonumber \\
    &\texttt{SelfConjugate   -> True,} \nonumber \\
	&\texttt{Indices         -> \{\},} \nonumber \\
	&\texttt{Mass            -> 2,} \nonumber \\
	&\texttt{MixingPartners  -> \{S[1], S[11]\}} \nonumber \\
    & \}
\end{align}

The couplings are defined in a straightforward way following the instructions given in the FeynArts manual. On expanding the summations in Eq.~(\ref{nsapsInt}), we find that there are twelve couplings or vertices for the NSAPS three-vector model. The four-point vertices corresponding to the couplings in the first line of Eq.~(\ref{nsapsInt}) are defined with a one component coupling constant. For instance, the four-point vertex corresponding to the term

$$ \frac{u_0}{6}\int_{q_1,...,q_4} \! \! \! \!\! \! \! \!\! \! \! \! \phitilde_1(q_1) \phi_1 (q_2) \phi_{1} (q_3) \phi_{1} (q_4) \, \deltabar(q_1 + q_2 +q_3+q_4)$$
is defined as 
\begin{equation}
\texttt{C[ S[1], S[1], S[1], S[11] ] == $\{\{$-u0$\}\}$},
\end{equation}
The three-point vertices corresponding to the couplings in the second line of Eq.~(\ref{nsapsInt}) are defined such that the parallel component of the momentum associated to the $\phitilde$ field in the coupling is taken into account. For this, we define each of the three point vertex with a two component coupling constant where only  one of the components is non-zero.  For example, the vertex corresponding to the interaction term

\begin{equation}
    i e_p  \int_{q_1,...,q_3} \! \! \! \!\! \! \! \!\! \! \! \! q_{1\|} \phitilde_1 (q_1) \phi_{2}(q_2) \phi_{3} (q_3) \deltabar (q_1+q_2+q_3)
\end{equation}
is defined as
\begin{align}
    &\texttt{C[ S[2] , S[3] , S[11] ] == }\nonumber \\
    &\texttt{\{\{-I*ep, 0\}, \{0, 0\}, \{0, 0\}\} }
\end{align}

The general form of the scalar and the mixed propagators are given in the generic file. FeynArts requires us to set the \textit{external} and the \textit{internal} propagator options separately. External (internal) propagator refers to the propagator corresponding to any of the external (internal) lines in a Feynman diagram. Since we are interested in the 1PI diagrams of the theory, for which the external lines are amputated, we set the external propagator to \texttt{1} by including the following line in the generic file:
\begin{align*}
    \texttt{AnalyticalPropagator[External]} \nonumber \\
    \texttt{[ s1 S[j1, mom] ] == 1}
\end{align*}
Note that we have not distinguished the set of $\phi$ fields from the set of $\widetilde{\phi}$ fields within the FeynArts files yet. FeynArts assumes that all the propagators corresponding to the various fields defined as scalars have the same form. However, we need the propagators corresponding to $\langle \widetilde{\phi}_i \widetilde{\phi}_i \rangle_0$ to vanish and the ones corresponding to $\langle \phi_i \phi_i \rangle_0$ to be of the form given in Eq.~(\ref{Gaussian_2pt}). This is done by introducing \texttt{Mass} as a factor in the form of the propagator as shown below.
\begin{align*}
&\texttt{AnalyticalPropagator[Internal][s1 S[j1, mom]] } \nonumber \\
&\texttt{== Mass[S[j1]]FAPropagatorDenominator} \nonumber \\
&\texttt{[mom,Mass[S[j1]]]},
\end{align*}
where \texttt{FAPropagatorDenominator} represents the explicit form of the scalar propagator, which is a function of the momentum \texttt{mom} and the attribute \texttt{Mass} of the corresponding scalar field \texttt{Mass[S[j1]]}. The explicit form can be given in the file \texttt{Alalytic.m} if need be. Since \texttt{Mass} is set  to \texttt{0} for the $\phitilde$ fields, the propagator vanishes whenever $\langle \phitilde_i \phitilde_i \rangle_0$ correlations are involved.  The form of the mixed propagator is given as
\begin{align*}
    & \texttt{AnalyticalPropagator[Internal][s1 Mix[S,S][j1,}  \nonumber \\ 
    & \texttt{mom]]==FAPropagatorDenominator[mom,} \nonumber \\
    & \texttt{Mass[Mix[S,S][j1]]]}.
\end{align*}
Note that the function \texttt{FAPropagatorDenominator} is used here for representing the mixed propagator.  The attribute \texttt{Mass} of the mixed and the scalar fields are used in the later part of the code, discussed below, for distinguishing the mixed from the scalar propagator.  

We make the following modifications in the file \texttt{Analytic.m} so that whenever the attribute \texttt{Mass} in the argument of the function \texttt{FAPropagatorDenominator} is \texttt{1} (\texttt{2}) the scalar (mixed) propagator is used.  

The original function 
\begin{align*}
\texttt{Format[FAPropagatorDenominator[p\_,m\_,d\_\_\_]] := } \\ 
\texttt{    Block[\{ x= p\textasciicircum2- m\^{}2 \}, 1/x\^{}d /; x=\!=0]} 
\end{align*}
is modified as 
\begin{align*}
\texttt{Format[FAPropagatorDenominator[p\_,m\_,d\_\_\_]] := } \\ 
\texttt{Block[\{x= PropNew[p\_,m\_,d\_\_\_]\}, 1/x\^{}d /; x=\!=0]}, 
\end{align*}
where \texttt{m} represents the attribute \texttt{Mass} and \texttt{p} the internal momentum. \texttt{PropNew} is an additional function written to the file as
\begin{align*}
    \texttt{PropNew[p\_,m\_,d\_\_\_]:=If[m==1,} \\
    \texttt{ScalarP@@\{p\}, MixP@@\{p\}]},
\end{align*}
 This is a conditional statement written in Wolfram language. In this calculation we do not use the explicit form of the propagator within the FeynArts environment. The functions \texttt{ScalarP} and \texttt{MixP} are used to represent the scalar and the mixed propagators, respectively, and are replaced by the explicit forms [see, Eq.~(\ref{Gaussian_2pt})] outside the FeynArts environment.

\subsection{Generating the diagrams and the expressions}
We now show how, having stored the information about the model in the classes and generic files, FeynArts is used to generate the required 1PI Feynman diagrams and the corresponding expressions.  
For this, we take the case of the NSAPS three-vector model and demonstrate, in particular,  how the one-loop contributions to $\Gamma_{1,1}^{11}(-p;p)$ can be obtained. This procedure can be extended to obtain  higher loop corrections to this and other vertex functions of this as well as other similar models.    
   
 The first step is to draw all the \textit{relevant} topologies. To identify these topologies we note that any Feynman diagram contributing to $\Gamma_{1,1}^{11}$ should have the following features.   
\begin{enumerate}
    \item It should be a 1PI diagram. 
    \item It should have two external lines, one each in the left and the right hand sides, corresponding to the external fields $\phi_1$ and $\phitilde_1$, respectively.  \item It should consist of only 3-point and 4-point vertices, since the model has only vertices of these kinds.
\end{enumerate}
It follows that the topologies that are relevant to the calculation of the one-loop corrections to $\Gamma_{1,1}^{11}$ are the ones that are irreducible, have two external lines, have only 3-nodes and 4-nodes, and have only one-loop.  All such topologies can be obtained using the \texttt{CreateTopologies} function provided by FeynArts as shown below.     \begin{align}
        \texttt{tops} =& \texttt{CreateTopologies[1, 1 -> 1,} \nonumber \\
        &\texttt{ExcludeTopologies -> Reducible]}
    \end{align}
This yields the following topologies. 
\begin{align} \label{1loop11top}
    \begin{tikzpicture}
    \begin{feynman}
    \vertex (a);
    \vertex[right=2cm of a] (b);
    \vertex[right=1cm of a] (c);
    \vertex[above=0.7cm of c] (o);
    \diagram*{
    (a) -- (b),
    (c)--[out=0, in=360, min distance =0.35cm](o),
    (c)--[out=180, in=180, min distance=0.35cm](o)
    };
    \end{feynman}
    \end{tikzpicture}
    \thickspace \; \; 
    \begin{tikzpicture}
    \begin{feynman}
    \vertex (a);
    \vertex[right=0.5cm of a] (b);
    \vertex[right=0.5cm of b] (i);
    \vertex[right=0.5cm of i] (c);
    \vertex[right=0.5cm of c] (d);
    \vertex[above=0.35cm of i] (o1);
    \vertex[below=0.35cm of i] (o2);
    \diagram*{
    (a) -- (b),
    (c) -- (d),
    (b) -- [out=40, in=180] (o1),
    (b) -- [out=320, in=180] (o2),
    (o1) --[out=0, in = 140] (c),
    (o2) --[out=0, in = 220] (c)
    };
    \end{feynman}
    \end{tikzpicture}
\end{align}

We now construct all possible realisations of these topologies that can be made using the building blocks of the theory (see Fig.~\ref{fig:buildingblocks}) such that the external lines are mixed propagators corresponding to $\langle \phi_1 \phitilde_1 \rangle_0$. These diagrams are rendered using the \texttt{InsertFields} function as shown below.
\begin{widetext}
\begin{align}
    \texttt{feynDiags =} &\texttt{ InsertFields[tops, Mix[S,S][11] -> Mix[S,S][11], InsertionLevel -> Classes, } \nonumber \\
    & \texttt{ ExcludeParticles->\{Mix[S,S][21], Mix[S,S][22], Mix[S,S][23]\},} \nonumber \\
    &\texttt{ Model -> FileNameJoin[\{\enquote{NSAPS}, \enquote{NSAPS}\}],GenericModel->NSAPSgen]}
\end{align}

This yields the following five Feynman diagrams. 
 
 \begin{align}
    \begin{tikzpicture}[baseline=(a.base)]
    \begin{feynman}[inline=(a.base)]
    \vertex (a);
    \vertex[right=2cm of a] (b);
    \vertex[right=1cm of a] (c);
    \vertex[above=0.7cm of c] (o) ;
    \vertex[below=0.05cm of c] (oa) {$\mathsmaller{1,0}$} ;
    \diagram*{
    (a) -- (b),
    (c)--[out=0, in=360, min distance =0.35cm, scalar](o),
    (c)--[out=180, in=180, min distance=0.35cm, scalar](o)
    };
    \end{feynman}
    \end{tikzpicture}
    \;
    \begin{tikzpicture}[baseline=(a.base)]
    \begin{feynman}[inline=(a.base)]
    \vertex (a);
    \vertex[right=2cm of a] (b);
    \vertex[right=1cm of a] (c);
    \vertex[above=0.7cm of c] (o) ;
    \vertex[below=0.05cm of c] (oa){$\mathsmaller{1,1}$};
    \diagram*{
    (a) -- (b),
    (c)--[out=0, in=360, min distance =0.35cm, scalar](o),
    (c)--[out=180, in=180, min distance=0.35cm, scalar](o)
    };
    \end{feynman}
    \end{tikzpicture}
    \;
    \begin{tikzpicture}[baseline=(a.base)]
    \begin{feynman}[inline=(a.base)]
    \vertex (a);
    \vertex[right=2cm of a] (b);
    \vertex[right=1cm of a] (c);
    \vertex[above=0.7cm of c] (o) ;
    \vertex[below=0.05cm of c] (oa) {$\mathsmaller{1,2}$} ;
    \diagram*{
    (a) -- (b),
    (c)--[out=0, in=360, min distance =0.35cm, scalar](o),
    (c)--[out=180, in=180, min distance=0.35cm, scalar](o)
    };
    \end{feynman}
    \end{tikzpicture} 
    \begin{tikzpicture}[baseline=(o2.base)]
    \begin{feynman}[inline=(o2.base)]
    \vertex (a);
    \vertex[right=0.5cm of a] (b);
    \vertex[right=0.5cm of b] (i);
    \vertex[right=0.5cm of i] (c);
    \vertex[right=0.5cm of c] (d);
    \vertex[above=0.35cm of i] (o1);
    \vertex[below=0.35cm of i] (o2);
    \vertex[below=0.05cm of b] (o1e) {$\mathsmaller{1}$} ;
    \vertex[below=0.05cm of c] (o2e) {$\mathsmaller{2}$} ;
    \diagram*{
    (a) -- (b),
    (c) -- (d),
    (b) -- [out=40, in=180] (o1),
    (b) -- [out=320, in=180, scalar] (o2),
    (o1) -- [out=0, in = 140] (c),
    (o2) -- [out=0, in = 220, scalar] (c)
    };
    \end{feynman}
    \end{tikzpicture}
    \;
    \begin{tikzpicture}[baseline=(o2.base)]
    \begin{feynman}[inline=(o2.base)]
    \vertex (a);
    \vertex[right=0.5cm of a] (b);
    \vertex[right=0.5cm of b] (i);
    \vertex[right=0.5cm of i] (c);
    \vertex[right=0.5cm of c] (d);
    \vertex[above=0.35cm of i] (o1);
    \vertex[below=0.35cm of i] (o2);
    \vertex[below=0.05cm of b] (o1e) {$\mathsmaller{1}$} ;
    \vertex[below=0.05cm of c] (o2e) {$\mathsmaller{3}$} ;
    \diagram*{
    (a) -- (b),
    (c) -- (d),
    (b) -- [out=40, in=180] (o1),
    (b) -- [out=320, in=180, scalar] (o2),
    (o1) -- [out=0, in = 140] (c),
    (o2) -- [out=0, in = 220, scalar] (c)
    };
    \end{feynman}
    \end{tikzpicture}
 \end{align}
\end{widetext}
The subscript below each 4-point vertex indicate the indices $a,i$ corresponding to the fields involved in the vertex (see Fig.~\ref{fig:buildingblocks}). Similarly, the subscript below each 3-point vertex indicate the corresponding index $a$. As is evident from the above, the topology
    \begin{tikzpicture}
    \begin{feynman}
    \vertex (a);
    \vertex[right=1cm of a] (b);
    \vertex[right=0.5cm of a] (c);
    \vertex[above=0.35cm of c] (o);
    \diagram*{
    (a) -- (b),
    (c)--[out=0, in=360, min distance =0.175cm](o),
    (c)--[out=180, in=180, min distance=0.175cm](o)
    };
    \end{feynman}
    \end{tikzpicture}
    can be realized using three of the nine four-point vertices shown in Fig.~\ref{fig:buildingblocks}. These are the ones with $a=1$ and $i=1$, $2$, and $3$, respectively. In all the three cases we obtain symmetrical Feynman diagrams.
    Similarly, the topology 
    \begin{tikzpicture}[baseline=(o2.base)]
    \begin{feynman}[inline=(o2.base)]
    \vertex (a);
    \vertex[right=0.25cm of a] (b);
    \vertex[right=0.25cm of b] (i);
    \vertex[right=0.25cm of i] (c);
    \vertex[right=0.25cm of c] (d);
    \vertex[above=0.175cm of i] (o1);
    \vertex[below=0.175cm of i] (o2);
    \diagram*{
    (a) -- (b),
    (c) -- (d),
    (b) -- [out=40, in=180] (o1),
    (b) -- [out=320, in=180] (o2),
    (o1) --[out=0, in = 140] (c),
    (o2) --[out=0, in = 220] (c)
    };
    \end{feynman}
    \end{tikzpicture}
    can be realised using two different pairs from the three three-point vertices given in Fig.~\ref{fig:buildingblocks}. The vertex corresponding to $a=1$ and $a=2$ make one pair and $a=1$ and $a=3$ make the other.
Now the function \texttt{CreateFeynAmp} is used to obtain the expressions corresponding to each diagram: 
    \begin{align}\label{oneloopcontribution}
    \begin{tikzpicture}[baseline=(a.base)]
    \begin{feynman}[inline=(a.base)]
    \vertex (a);
    \vertex[right=2cm of a] (b);
    \vertex[right=1cm of a] (c);
    \vertex[above=0.7cm of c] (o) ;
    \vertex[below=0.05cm of c] (oa) {$\mathsmaller{1,0}$} ;
    \diagram*{
    (a) -- (b),
    (c)--[out=0, in=360, min distance =0.35cm, scalar](o),
    (c)--[out=180, in=180, min distance=0.35cm, scalar](o)
    };
    \end{feynman}
    \end{tikzpicture}
    &= -\frac{u_0}{2} \int_{q_1} C_0 (q_1),
    \end{align}
    \begin{align}
    \begin{tikzpicture}[baseline=(a.base)]
    \begin{feynman}[inline=(a.base)]
    \vertex (a);
    \vertex[right=2cm of a] (b);
    \vertex[right=1cm of a] (c);
    \vertex[above=0.7cm of c] (o) ;
    \vertex[below=0.05cm of c] (oa) {$\mathsmaller{1,1}$} ;
    \diagram*{
    (a) -- (b),
    (c)--[out=0, in=360, min distance =0.35cm, scalar](o),
    (c)--[out=180, in=180, min distance=0.35cm, scalar](o)
    };
    \end{feynman}
    \end{tikzpicture}
    =
    \begin{tikzpicture}[baseline=(a.base)]
    \begin{feynman}[inline=(a.base)]
    \vertex (a);
    \vertex[right=2cm of a] (b);
    \vertex[right=1cm of a] (c);
    \vertex[above=0.7cm of c] (o) ;
    \vertex[below=0.05cm of c] (oa) {$\mathsmaller{1,2}$} ;
    \diagram*{
    (a) -- (b),
    (c)--[out=0, in=360, min distance =0.35cm, scalar](o),
    (c)--[out=180, in=180, min distance=0.35cm, scalar](o)
    };
    \end{feynman}
    \end{tikzpicture}
    &= - \frac{u_1}{2} \int_{q_1} C_0(q_1)
\end{align}
\begin{align}
    \begin{tikzpicture}[baseline=(a.base)]
    \begin{feynman}[inline=(a.base)]
    \vertex (a);
    \vertex[right=0.5cm of a] (b);
    \vertex[right=0.5cm of b] (i);
    \vertex[right=0.5cm of i] (c);
    \vertex[right=0.5cm of c] (d);
    \vertex[above=0.35cm of i] (o1);
    \vertex[below=0.35cm of i] (o2);
    \vertex[below=0.05cm of b] (o1e) {$\mathsmaller{1}$};
    \vertex[below=0.05cm of c] (o2e) {$\mathsmaller{2}$};
    \diagram*{
    (a) -- (b),
    (c) -- (d),
    (b) -- [out=40, in=180] (o1),
    (b) -- [out=320, in=180, scalar] (o2),
    (o1) -- [out=0, in = 140] (c),
    (o2) -- [out=0, in = 220, scalar] (c)
    };
    \end{feynman}
    \end{tikzpicture}
    =
    \begin{tikzpicture}[baseline=(a.base)]
    \begin{feynman}[inline=(a.base)]
    \vertex (a);
    \vertex[right=0.5cm of a] (b);
    \vertex[right=0.5cm of b] (i);
    \vertex[right=0.5cm of i] (c);
    \vertex[right=0.5cm of c] (d);
    \vertex[above=0.35cm of i] (o1);
    \vertex[below=0.35cm of i] (o2);
    \vertex[below=0.05cm of b] (o1e) {$\mathsmaller{1}$};
    \vertex[below=0.05cm of c] (o2e) {$\mathsmaller{3}$};
    \diagram*{
    (a) -- (b),
    (c) -- (d),
    (b) -- [out=40, in=180] (o1),
    (b) -- [out=320, in=180, scalar] (o2),
    (o1) -- [out=0, in = 140] (c),
    (o2) -- [out=0, in = 220, scalar] (c)
    };
    \end{feynman}
    \end{tikzpicture}
    &= -e_p^2 \int_{q_1} p_\| (p_\|-q_{1\|}) \nonumber \\
    &\text{ }\; \; \times C_0(q_1) G_0(p-q_1)
    \end{align}

In the rest of this paper symmetrical Feynman diagrams constructed with different combinations of the vertices will not be shown separately. Instead, only a single diagram without any subscripts will be used and is to be understood as the sum of all the symmetrical diagrams. For example, the diagrams 
\begin{widetext}
\begin{align}\label{oneloopcon}
 &\begin{tikzpicture}
    \begin{feynman}
    \vertex (i);
    \vertex[below=0.5cm of i] (a);
    \vertex[right=2cm of a] (b);
    \vertex[right=1cm of a] (c);
    \vertex[above=0.7cm of c] (o) ;
    \vertex[above=0.05cm of o] (oa) ;
    \diagram*{
    (a) -- (b),
    (c)--[out=0, in=360, min distance =0.35cm, scalar](o),
    (c)--[out=180, in=180, min distance=0.35cm, scalar](o)
    };
    \end{feynman}
    \end{tikzpicture}   
    =
    \begin{tikzpicture}[baseline=(a.base)]
    \begin{feynman}[inline=(a.base)]
    \vertex (a);
    \vertex[right=2cm of a] (b);
    \vertex[right=1cm of a] (c);
    \vertex[above=0.7cm of c] (o) ;
    \vertex[below=0.05cm of c] (oa) {$\mathsmaller{1,0}$} ;
    \diagram*{
    (a) -- (b),
    (c)--[out=0, in=360, min distance =0.35cm, scalar](o),
    (c)--[out=180, in=180, min distance=0.35cm, scalar](o)
    };
    \end{feynman}
    \end{tikzpicture}  
    +
    \begin{tikzpicture}[baseline=(a.base)]
    \begin{feynman}[inline=(a.base)]
    \vertex (a);
    \vertex[right=2cm of a] (b);
    \vertex[right=1cm of a] (c);
    \vertex[above=0.7cm of c] (o) ;
    \vertex[below=0.05cm of c] (oa) {$\mathsmaller{1,1}$} ;
    \diagram*{
    (a) -- (b),
    (c)--[out=0, in=360, min distance =0.35cm, scalar](o),
    (c)--[out=180, in=180, min distance=0.35cm, scalar](o)
    };
    \end{feynman}
    \end{tikzpicture}
    +
    \begin{tikzpicture}[baseline=(a.base)]
    \begin{feynman}[inline=(a.base)]
    \vertex (a);
    \vertex[right=2cm of a] (b);
    \vertex[right=1cm of a] (c);
    \vertex[above=0.7cm of c] (o) ;
    \vertex[below=0.05cm of c] (oa) {$\mathsmaller{1,2}$} ;
    \diagram*{
    (a) -- (b),
    (c)--[out=0, in=360, min distance =0.35cm, scalar](o),
    (c)--[out=180, in=180, min distance=0.35cm, scalar](o)
    };
    \end{feynman}
    \end{tikzpicture}
    = -\left(\frac{u_0}{2} + u_1 \right) \int_{q_1} C_0 (q_1), \; \text{and} \nonumber \\
    &\begin{tikzpicture}[baseline=(a.base)]
    \begin{feynman}[inline=(a.base)]
    \vertex (a);
    \vertex[right=0.5cm of a] (b);
    \vertex[right=0.5cm of b] (i);
    \vertex[right=0.5cm of i] (c);
    \vertex[right=0.5cm of c] (d);
    \vertex[above=0.35cm of i] (o1);
    \vertex[below=0.35cm of i] (o2);
    \diagram*{
    (a) -- (b),
    (c) -- (d),
    (b) -- [out=40, in=180] (o1),
    (b) -- [out=320, in=180, scalar] (o2),
    (o1) -- [out=0, in = 140] (c),
    (o2) -- [out=0, in = 220, scalar] (c)
    };
    \end{feynman}
    \end{tikzpicture}
    =
    \begin{tikzpicture}[baseline=(a.base)]
    \begin{feynman}[inline=(a.base)]
    \vertex (a);
    \vertex[right=0.5cm of a] (b);
    \vertex[right=0.5cm of b] (i);
    \vertex[right=0.5cm of i] (c);
    \vertex[right=0.5cm of c] (d);
    \vertex[above=0.35cm of i] (o1);
    \vertex[below=0.35cm of i] (o2);
    \vertex[below=0.05cm of b] (o1e) {$\mathsmaller{1}$};
    \vertex[below=0.05cm of c] (o2e) {$\mathsmaller{2}$};
    \diagram*{
    (a) -- (b),
    (c) -- (d),
    (b) -- [out=40, in=180] (o1),
    (b) -- [out=320, in=180, scalar] (o2),
    (o1) -- [out=0, in = 140] (c),
    (o2) -- [out=0, in = 220, scalar] (c)
    };
    \end{feynman}
    \end{tikzpicture}
    +
    \begin{tikzpicture}[baseline=(a.base)]
    \begin{feynman}[inline=(a.base)]
    \vertex (a);
    \vertex[right=0.5cm of a] (b);
    \vertex[right=0.5cm of b] (i);
    \vertex[right=0.5cm of i] (c);
    \vertex[right=0.5cm of c] (d);
    \vertex[above=0.35cm of i] (o1);
    \vertex[below=0.35cm of i] (o2);
    \vertex[below=0.05cm of b] (o1e) {$\mathsmaller{1}$};
    \vertex[below=0.05cm of c] (o2e) {$\mathsmaller{3}$};
    \diagram*{
    (a) -- (b),
    (c) -- (d),
    (b) -- [out=40, in=180] (o1),
    (b) -- [out=320, in=180, scalar] (o2),
    (o1) -- [out=0, in = 140] (c),
    (o2) -- [out=0, in = 220, scalar] (c)
    };
    \end{feynman}
    \end{tikzpicture}
    = -2 e_p^2 \int_{q_1} p_\| (p_\|-q_{1\|}) C_0(q_1) G_0(p-q_1)
\end{align}
Using Eq.~(\ref{oneloopcon}), we write
\begin{align}\label{gamma11oneloop}
    \Gamma_{1,1}^{11}(p;-p)\;= \; 
ip_0+M(\boldsymbol{p})
-
\underbrace{
\left[
\begin{tikzpicture}
    \begin{feynman}
    \vertex (i);
    \vertex[below=0.5cm of i] (a);
    \vertex[right=2cm of a] (b);
    \vertex[right=1cm of a] (c);
    \vertex[above=0.7cm of c] (o) ;
    \vertex[above=0.05cm of o] (oa) ;
    \diagram*{
    (a) -- (b),
    (c)--[out=0, in=360, min distance =0.35cm, scalar](o),
    (c)--[out=180, in=180, min distance=0.35cm, scalar](o)
    };
    \end{feynman}
    \end{tikzpicture}
+ 
\begin{tikzpicture}
    \begin{feynman}
    \vertex (a);
    \vertex[right=0.5cm of a] (b);
    \vertex[right=0.5cm of b] (i);
    \vertex[right=0.5cm of i] (c);
    \vertex[right=0.5cm of c] (d);
    \vertex[above=0.35cm of i] (o1);
    \vertex[below=0.35cm of i] (o2);
    \diagram*{
    (a) -- (b),
    (c) -- (d),
    (b) -- [out=40, in=180] (o1),
    (b) -- [out=320, in=180, scalar] (o2),
    (o1) -- [out=0, in = 140] (c),
    (o2) -- [out=0, in = 220, scalar] (c)
    };
    \end{feynman}
    \end{tikzpicture}
    \right]
    }_{\Sigma_1 (p)} + 
    \text{ higher loop corrections}
\end{align}

Divergent integrals appear at one and higher loops. In the next section, we use the one-loop corrections to $\Gamma_{1,1}^{11}$, denoted by $\Sigma_1$ in Eq.~(\ref{gamma11oneloop}), as a demonstrative example for explaining the procedure for regularizing the divergent Feynman integrals. 
Using Eq.~(\ref{oneloopcon}) and Eq.~(\ref{gamma11oneloop}) we explicitly write 
\begin{align}\label{sigma1}
    \Sigma_1 (p)= -\left(\frac{u_0}{2} + u_1 \right) \int_{q_1} C_0 (q_1) -2 e_p^2 \int_{q_1} p_\| (p_\|-q_{1\|}) C_0(q_1) G_0(p-q_1).
\end{align}
\end{widetext}
\section{Dimensional regularization of the UV divergent integrals} \label{sec:dimreg}
The Feynman diagrams appearing in the NSAPS three-vector model as well as other similar statistical models, which may be UV divergent, yield integrals that have certain similarities. These similarities are exploited to formulate a general procedure to dimensional regularize the integrals. In this section we explain this procedure, which can be easily automatized using Mathematica with the help of the packages SecDec and Piecewise, in the context of the NSAPS three-vector model.

Once all the diagrams contributing to the relevant vertex functions up to the desired loop and the corresponding expressions are obtained, we may isolate the divergences. For example, the UV divergences lurking in $\Gamma_{1,1}^{11}$ can be isolated by considering the quantities 
\begin{align}
    \Gamma_{1,1}^{11}(0), \;  \frac{\partial^2 }{\partial p_\|^2}\Gamma_{1,1}^{11} (0), \;  \frac{\partial^2 }{\partial p_\perp^2}\Gamma_{1,1}^{11} (0), \; \text{and} \; \frac{\partial^2 }{\partial i p_0}\Gamma_{1,1}^{11} (0)
\end{align}
At one-loop we can equivalently consider the following quantities. 
\begin{align}
        \Sigma_1 (0) =-\left(\frac{u_0}{2} + u_1 \right) \int_{q_1} C_0 (q_1). 
\end{align}
\begin{align}
     \frac{\partial^2 }{\partial p_\|^2} \Sigma_1(0) =& 8 D T \rho \, e_p^2 \int_{q_1} q_{1\|}^2 G_0 (q_1)^2 \widetilde{C}_0 (q_1) \nonumber \\
     &- 4 T e_p^2 \int_{q_1} G_0 (q_1) \widetilde{C}_0 (q_1)  \label{sigma12}
\end{align}
\begin{align}
        \frac{\partial^2 }{\partial p_\perp^2} \Sigma_1(0)=0, \; \frac{\partial }{\partial ip_0} \Sigma_1(0)=0.
\end{align}
where we have used Eq.~(\ref{Gaussian_2pt}) and (\ref{sigma1}) and introduced the rescaled scalar propagator
\begin{align}
    \widetilde{C}_0 (q_1) = \frac{C_0}{2T}.
\end{align}
Note that only $\Sigma_1 (0) $ and $     \frac{\partial^2 }{\partial p_\|^2} \Sigma_1(0)$ are non-zero, the integrals wherein are UV divergent. Such integrals appearing in various vertex functions at various loops are in general of the form 
\begin{align}\label{genform1}
    \mathcal{J}=\int_{q_1,...,q_l} &\Big{[} (q_{1\alpha})^{m_{\alpha_1}}
    (q_{1\beta})^{m_{\beta_1}}...(q_{2\alpha})^{m_{\alpha_2}}
    (q_{2\beta})^{m_{\beta_2}}...\Big{]} \nonumber \\ \times&\left[G_0(\widetilde{q}_1)...G_0(\widetilde{q}_{n_1}) \widetilde{C}_0 (\widetilde{q}_{n_1+1})...\widetilde{C}_0 (\widetilde{q}_{n_1+n_2}) \right],
\end{align}
where each $\widetilde{q}_i$ is a linear combination of the internal momenta $q_i$ and $l$ is the loop order. The Greek indices that are written as subscripts to the internal momenta indicate their various components, and the powers $m_{\alpha_i},\, m_{\beta_i},$ etc. are integers.

Before performing dimensional regularization and expanding the divergent parts of the integral in Eq.~(\ref{genform1}) in negative powers of $\epsilon=d-d_c$, we first integrate out the time conjugate  variables $q_{10},...q_{l0}$. In the integral considered in Eq.~(\ref{genform1}) only the propagator part depends on the time conjugate  variables. This is in general the case at any loop for any vertex function if the propagators of the theory are either of the form  
\begin{align}
    F(q) \propto \frac{1}{q_0 + f(\boldsymbol{q})} 
\end{align}
or a product of functions of the above form and if the vertex momenta are independent of the time component. This is indeed the case for the NSAPS three-vector model as is evident from Eqs.~(\ref{Gaussian_2pt}) and (\ref{nsapsInt}). Note that the scalar propagator can be written as $C_0(q)=2 T G_0(q) G_0(-q).$ 
It follows that we can write Eq.~(\ref{genform1}) as
\begin{align}\label{genform2}
    \mathcal{J}=\int_{\boldsymbol{q}_1,...,\boldsymbol{q}_l} \Big{[} (q_{1\alpha})^{m_{\alpha_1}}
    (q_{1\beta})^{m_{\beta_1}}...(q_{2\alpha})^{m_{\alpha_2}}
    (q_{2\beta})^{m_{\beta_2}}...\Big{]}\, \mathcal{I},
\end{align}
where
\begin{align}\label{stepint1}
   \mathcal{I}= \int_{q_{10}...q_{l0}}     G_0(\widetilde{q}_1)...G_0(\widetilde{q}_{n_1}) \widetilde{C}_0 (\widetilde{q}_1)...\widetilde{C}_0 (\widetilde{q}_{n_2}).
\end{align}
To evaluate the integral in Eq.~(\ref{stepint1}), we write $G_0$ and $C_0$ in terms of their temporal Fourier counterparts: 
\begin{align}\label{fourierG0}
    G_0(\widetilde{q}_{0}, \widetilde{\boldsymbol{q}}) &= \int_t G_0(t, \widetilde{\boldsymbol{q}}) \exp{(i \widetilde{q}_{0} t)}, \nonumber \\ \widetilde{C}_0(\widetilde{q}_{0}, \widetilde{\boldsymbol{q}}) &= \int_t \widetilde{C}_0(t, \widetilde{\boldsymbol{q}}) \exp{(i \widetilde{q}_{0} t)}.
\end{align}
where
\begin{align}\label{propsintime}
    G_0(t,\widetilde{\boldsymbol{q}})=e^{-M(\widetilde{\boldsymbol{q}})t} \, \Theta(t),  \; \; \widetilde{C}_0 (t, \widetilde{q}) = \frac{1}{2 M(\widetilde{\boldsymbol{q}})}e^{- M(\widetilde{\boldsymbol{q}})|t| }
\end{align}
\begin{align}
    \Theta(t)=
    \begin{cases}
    1, & t \ge 0, \\
    0, & t <0, 
    \end{cases}
    \;  \; M. (\widetilde{\boldsymbol{q}})=D ( \widetilde{\boldsymbol{q}}_\perp^2 + \rho q_\|^2 +r), 
\end{align}
and $\int_t \equiv \int dt$. The temporal Fourier transforms $G_0(t,\widetilde{\boldsymbol{q}})$ and $\widetilde{C}_0(t,\widetilde{\boldsymbol{q}})$ are obtained from Eq.~(\ref{Gaussian_2pt}). Using Eq.~(\ref{fourierG0}) in Eq.~(\ref{stepint1}) we get 
\begin{widetext}
\begin{align}\label{stepint2}
\mathcal{I}=
        \int_{q_{10}...q_{l0}} \int_{t_1,...t_{n_1+n_2}}  \! \!\!\! \! \!\!\!  \! \!\!\! G_0 (t_1,\widetilde{q}_1)...G_0(t_{n_1},\widetilde{\boldsymbol{q}}_{n_1})\mathcal{C}_0 (t_{n_1+1},\widetilde{\boldsymbol{q}}_{n_1+1})...\mathcal{C}_0(t_{n_1+n_2},\widetilde{\boldsymbol{q}}_{n_1+n_2})
        \exp{ \left\{i \left(\underbrace{\widetilde{q}_{10}t_{1}+...+\widetilde{q}_{n_1+n_20} t_{n_1+n_2}}_{D}\right) \right\}}
\end{align}
\end{widetext}
Since $\widetilde{q_i}$ are linear combinations of $q_{i}$, the sum appearing in the exponent above, denoted by $D$, can be written in the form
\begin{align}
    D=T_1 q_{10}+...+T_l q_{l0},
\end{align}
where $T_i$ are linear combinations of $t_i$. Clearly, we can integrate out $q_{i0}$ from Eq.~(\ref{stepint2}) to obtain a product of $l$ delta functions yielding
\begin{align}\label{instep1}
\mathcal{I}=&\int_{t_1,...t_{n_1+n_2}}  \! \!\!\! \! \!\!\!  \! \!\!\! G_0 (t_1,\widetilde{q}_1)...G_0(t_{n_1},\widetilde{\boldsymbol{q}}_{n_1})\nonumber \\
&\mathcal{C}_0 (t_{n_1+1},\widetilde{\boldsymbol{q}}_{n_1+1})...\mathcal{C}_0(t_{n_1+n_2},\widetilde{\boldsymbol{q}}_{n_1+n_2})\, \delta(T_1)...\delta(T_{l}).    
\end{align}
After substituting for $G_0$ and $\widetilde{C}_0$ in Eq.~(\ref{instep1}) using Eq.~(\ref{propsintime}), we can easily integrate out the time variables $t_1,..t_{n_1+n_2}$. This can be implemented using the Integrate function in Mathematica or the package Piecewise. Once the time conjugate variables $q_{i0}$ are integrated out, we substitute $\mathcal{I}$ back in Eq.~(\ref{genform2}) and rescale the components of internal momenta as $\{q_{i\|},\boldsymbol{q}_{i\perp\}} \rightarrow \{\sqrt{\rho} q_\|, \sqrt{\rho/r} \boldsymbol{q}_{i\perp} \} $ to yield $\mathcal{J}$ in the form 
\begin{align}\label{genform3}
    \mathcal{J}=& C \int_{\boldsymbol{q}_1...\boldsymbol{q}_l} \frac{ (q_{1\alpha})^{m_{\alpha_1}}
    (q_{1\beta})^{m_{\beta_1}}...(q_{2\alpha})^{m_{\alpha_2}}
    (q_{2\beta})^{m_{\beta_2}}...}{\mathcal{M}_1\mathcal{M}_2...} \nonumber \\
    &+ \text{ similar terms},
\end{align}
where each $\mathcal{M}_i$ is a linear combination of $M(\boldsymbol{\widetilde{q}}_i) \equiv \boldsymbol{\widetilde{q}}_i^2+1$, $\widetilde{q}_i$ represents a linear combination of the rescaled internal momenta $q_i$, and $C$ is a constant factor containing the parameters of the theory.  

For demonstrative purposes, we explain the case of $\Sigma_{1\|} \equiv \frac{\partial^2 }{\partial p_\|^2} \Sigma_1(0)$ given in Eq.~(\ref{sigma1}). First, we separate the integration with respect $q_{10}$ to obtain
\begin{align}\label{instep10}
 \Sigma_{1\|}=&   
 8 D T \rho \, e_p^2 \int_{\boldsymbol{q}_1} q_{1\|}^2 \int_{q_{10}}G_0 (q_1)^2 \widetilde{C}_0 (q_1) \nonumber \\
 &- 4 T e_p^2 \int_{\boldsymbol{q}_1} \int_{q_{10}}G_0 (q_1) \widetilde{C}_0 (q_1)
\end{align}
The $q_{10}$ integral in the first term is evaluated as
\begin{align}\label{q10int1}
    &\int_{q_{10}}G_0 (q_1)^2 \widetilde{C}_0 (q_1) \nonumber \\
    &= \int_{q_{10}} \int_{t_1, t_2, t_3} \widetilde{C}_0(t_1,\boldsymbol{q}_1) G_0 (t_2,\boldsymbol{q}_1) G_0(t_3,\boldsymbol{q}_1) \nonumber \\
    &\; \; \; \; \exp{\{ i q_{10} (t_1+t_2+t_3) \}} \nonumber \\
    &= \int_{t_1, t_2, t_3} \widetilde{C}_0(t_1,\boldsymbol{q}_1) G_0 (t_2,\boldsymbol{q}_1) G_0(t_3,\boldsymbol{q}_1) \delta(t_1 +t_2+t_3) \nonumber \\
    &=\int_{t_1, t_2, t_3} \frac{e^{ -(|t_1|+t_2+t_3) M(\boldsymbol{q}_1)}}{2 M(\boldsymbol{q}_1)} \Theta(t_2) \Theta(t_3) \delta(t_1 +t_2+t_3) \nonumber \\
    &=\frac{1}{8 M(\boldsymbol{q}_1)^3}
\end{align}
In the one to last step we have used Eq.~(\ref{fourierG0}).  Proceeding in similar fashion, we obtain
\begin{align}\label{q10int2}
    \int_{q_{10}}G_0 (q_1) \widetilde{C}_0 (q_1) = \frac{1}{4 M(\boldsymbol{q}_1)^2}
\end{align}
Substituting for the $q_{10}$ integrals in Eq.~(\ref{instep10}) using Eqs.~(\ref{q10int1}) and (\ref{q10int2}), we obtain
\begin{equation}
 \Sigma_{1\|}=   DT \rho e_p^2 \int_{\boldsymbol{q}_1}\frac{q_{1\|}^2}{M(\boldsymbol{q}_1)^3} -T e_p^2 \int_{\boldsymbol{q}_1} \frac{1}{M(\boldsymbol{q}_1)^2}
\end{equation}
After rescaling the components of the internal momenta as $\{q_\|,\boldsymbol{q}_\perp\} \rightarrow \{\sqrt{\rho} q_\|, \sqrt{\rho/r} \boldsymbol{q}_\perp \} $, we write 
\begin{align}\label{aftTConInt}
    \Sigma_{1\|}= \frac{Tr^{(d-1)/2}}{D^2 \sqrt{\rho}} e_p^2 \int_{\boldsymbol{q}_1} \frac{q_{1\|}^2}{M(\boldsymbol{q}_1)^3} - 
    \frac{Tr^{(d-1)/2}}{D^2 \sqrt{\rho}} e_p^2 \int_{\boldsymbol{q}_1} \frac{1}{M(\boldsymbol{q}_1)^2}.
\end{align}
Now, consider the integral in Eq.~(\ref{genform3}): 
\begin{align}\label{tensorfeynman}
    \mathcal{J}_1= \int_{\boldsymbol{q}_1...\boldsymbol{q}_l} \frac{ (q_{1\alpha})^{m_{\alpha_1}}
    (q_{1\beta})^{m_{\beta_1}}...(q_{2\alpha})^{m_{\alpha_2}}
    (q_{2\beta})^{m_{\beta_2}}...}{\mathcal{M}_1\mathcal{M}_2...},
\end{align}
which is a tensor Feynman integral. To use computational packages like SecDec for numerical dimensional regularization the tensor Feynman integrals should be expressed in terms of scalar integrals. This is especially simple in the case of the integrals in Eq.~(\ref{tensorfeynman}) as there is no dependence on the external momenta. Examples of two kinds of tensor Feynman integrals that appear in the NSAPS three-vector model are shown below. 

\begin{enumerate}
\item 
\begin{align}
  \int_{\boldsymbol{q}_1,... \boldsymbol{q}_l}  \frac{q_{i \alpha} q_{j\beta} }{\mathcal{M}_1 \mathcal{M}_2 ...}=\delta_{\alpha \beta} I_0.
\end{align}
Contracting with $\delta^{\alpha \beta}$ on both sides we obtain
\begin{align}\label{tensorFyn1}
   I_0= \frac{1}{4}     \int_{\boldsymbol{q}_1,... \boldsymbol{q}_l}  \frac{\boldsymbol{q}_{i} .\boldsymbol{q}_{j} }{\mathcal{M}_1 \mathcal{M}_2 ...}
\end{align}
\item \begin{align}
    \int_{\boldsymbol{q}_1...\boldsymbol{q}_l} \frac{q_{i \alpha} q_{i\beta} q_{j \gamma} q_{j\delta}}{\mathcal{M}_1 \mathcal{M}_2 ...} = \delta_{\alpha \beta} \delta_{\gamma \delta} I_1 + \delta_{\alpha \gamma} \delta_{\beta \delta} I_2 + \delta_{\alpha \delta} \delta_{\beta \gamma} I_3
\end{align}
Contracting both the left and the right hand sides with $\delta_{\alpha \beta} \delta_{\gamma \delta} $, $\delta_{\alpha \gamma} \delta_{\beta \delta}$ and $\delta_{\alpha \delta} \delta_{\beta \gamma}$ and then solving for $I_1$, $I_2$ and $I_3$ yields  
\begin{align}
    I_1=&
    -\frac{(d+1)}{d(d-1)(2+d)} \int_{\boldsymbol{q}_1...\boldsymbol{q}_l} \frac{\boldsymbol{q}_i^2 \boldsymbol{q}_j^2}{\mathcal{M}_1 \mathcal{M}_2 ...} \nonumber \\
    &+\frac{2}{d(d-1)(d+2)} \int_{\boldsymbol{q}_1...\boldsymbol{q}_l} \frac{(\boldsymbol{q}_i. \boldsymbol{q}_j)^2}{\mathcal{M}_1 \mathcal{M}_2 ...} \nonumber \\
    I_2 =I_3 =&    - \frac{1}{d(d-1)(d+2)} \int_{\boldsymbol{q}_1...\boldsymbol{q}_l} \frac{\boldsymbol{q}_i^2 \boldsymbol{q}_j^2}{\mathcal{M}_1 \mathcal{M}_2 ...} \nonumber \\
    &+\frac{d}{d(d-1)(d+2)} \int_{\boldsymbol{q}_1...\boldsymbol{q}_l} \frac{(\boldsymbol{q}_i. \boldsymbol{q}_j)^2}{\mathcal{M}_1 \mathcal{M}_2 ...}
\end{align}
\end{enumerate}
The scalar Feynman integrals can be computed by first Feynman parameterizing them and then integrating out the internal momenta followed by numerically evaluating the parameter integrals by sector decomposition. Several computational packages are available that can perform these tasks. We use the package SecDec which can be used with Mathematica. 

Consider, for example, the Feynman integrals appearing in $\Sigma_{1\|}$ given in Eq.~(\ref{aftTConInt}). The tensor Feynman integral appearing in the first term therein can be written in terms of a scalar integral as in Eq.~(\ref{tensorFyn1}):   
\begin{align}\label{tensorinsigma}
    \int_{\boldsymbol{q}_1}\frac{q_{1\|}^2}{M(\boldsymbol{q}_1)^3}=\frac{1}{4}\int_{\boldsymbol{q}_1}\frac{\boldsymbol{q}_{1}^2}{M(\boldsymbol{q}_1)^3}
\end{align}
The integral in the right hand side of Eq.~(\ref{tensorinsigma}) can be regularized numerically using SecDec to yield
\begin{align}\label{tensorinsigma2}
    \int_{\boldsymbol{q}_1}\frac{\boldsymbol{q}_{1}^2}{M(\boldsymbol{q}_1)^3}= \frac{1}{64 \pi^2 \epsilon }+ \text{ regular terms}, 
\end{align}
where $\epsilon=4-d$ and the fact that the critical dimension $d_c=4$ for the NSAPS three-vector model is used. 
Similarly, the scalar integral in the second term in the right hand side of Eq.~(\ref{aftTConInt}) is evaluated as
\begin{equation}\label{scalarinsigma}
    \int_{\boldsymbol{q}_1} \frac{1}{M(\boldsymbol{q}_1)^2}=\frac{1}{16 \pi^2 \epsilon} + \text{  regular terms}
\end{equation}
Using Eqs.~(\ref{tensorinsigma}), (\ref{tensorinsigma2}), and (\ref{scalarinsigma}) in Eq.~(\ref{aftTConInt}), we obtain
\begin{equation}
\Sigma_{1 \|}
=- \frac{0.75 Tr^{(d-1)/2}}{16 \pi^2 D^2 \sqrt{\rho}} e_p^2  \left( \frac{1}{\epsilon } + \text{regular terms} \right).
\end{equation}
To summarize, we showed how FeynArts can be used to generate the 1PI Feynman diagrams for a statistical field theory using the example of the NSAPS three-vector model. We then presented a procedure, which can be easily automatized in Mathematica, for simplifying the integrals obtained from the diagrams so that they can be regularized using SecDec. We demonstrated the former part by generating the one-loop 1PI diagrams contributing to $\Gamma_{1,1}^{11}$, and the latter part by regularizing the UV divergent integrals appearing in $\Sigma_{1\|}$. As already stated, the various procedures discussed in this work can be generalized to be applied to other similar MSR field theories. 
\bibliographystyle{unsrt}
\bibliography{References2}

\begin{thebibliography}{10}

\bibitem{polchinski_1998}
Joseph Polchinski.
\newblock {\em String Theory}, volume~1 of {\em Cambridge Monographs on
  Mathematical Physics}.
\newblock Cambridge University Press, 1998.

\bibitem{Peskin_2018}
Michael~E Peskin.
\newblock {\em An introduction to quantum field theory}.
\newblock CRC press, 2018.

\bibitem{tauber_2014}
Uwe~C. Täuber.
\newblock {\em Critical Dynamics: A Field Theory Approach to Equilibrium and
  Non-Equilibrium Scaling Behavior}.
\newblock Cambridge University Press, 2014.

\bibitem{Halperin_1977}
P.~C. Hohenberg and B.~I. Halperin.
\newblock Theory of dynamic critical phenomena.
\newblock {\em Rev. Mod. Phys.}, 49:435--479, 1977.

\bibitem{Martin_1973}
Paul~Cecil Martin, ED~Siggia, and HA~Rose.
\newblock {\em Physical Review A}, 8(1):423, 1973.

\bibitem{Zinn_2002}
Jean Zinn-Justin.
\newblock {\em {Quantum Field Theory and Critical Phenomena}}.
\newblock Oxford University Press, 2002.

\bibitem{Janssen_1986}
HK~Janssen and B~Schmittmann.
\newblock Field theory of critical behaviour in driven diffusive systems.
\newblock {\em Z. Phys. B}, 64(4):503--514, 1986.

\bibitem{Janssen_1981}
Hans-Karl Janssen.
\newblock On the nonequilibrium phase transition in reaction-diffusion systems
  with an absorbing stationary state.
\newblock {\em Z. Phys. B}, 42(2):151--154, 1981.

\bibitem{Bassler_1994}
KE~Bassler and Beate Schmittmann.
\newblock Renormalization-group study of a hybrid driven diffusive system.
\newblock {\em Phys. Rev. E}, 49(5):3614, 1994.

\bibitem{Schmittmann_1995}
B.~Schmittmann and R.K.P. Zia.
\newblock Phase transitions and critical phenomena.
\newblock In C.~Domb and J.~Lebowitz, editors, {\em Phase Transitions and
  Critical Phenomena}, volume~17. Academic Press, London, 1995.

\bibitem{feynarts2001}
Thomas Hahn.
\newblock Generating feynman diagrams and amplitudes with feynarts 3.
\newblock {\em Computer Physics Communications}, 140(3):418--431, 2001.

\bibitem{qgraph1993}
Paulo Nogueira.
\newblock Automatic feynman graph generation.
\newblock {\em Journal of Computational Physics}, 105(2):279--289, 1993.

\bibitem{secdec2015}
Sophia Borowka, Gudrun Heinrich, SP~Jones, M~Kerner, J~Schlenk, and T~Zirke.
\newblock Secdec-3.0: numerical evaluation of multi-scale integrals beyond one
  loop.
\newblock {\em Computer Physics Communications}, 196:470--491, 2015.

\bibitem{fiesta2009}
AV~Smirnov and MN~Tentyukov.
\newblock Feynman integral evaluation by a sector decomposition approach
  (fiesta).
\newblock {\em Computer Physics Communications}, 180(5):735--746, 2009.

\bibitem{feyncalc2016}
Vladyslav Shtabovenko, Rolf Mertig, and Frederik Orellana.
\newblock New developments in feyncalc 9.0.
\newblock {\em Computer Physics Communications}, 207:432--444, 2016.

\bibitem{fare2016}
Michele Re~Fiorentin.
\newblock Fare: a mathematica package for tensor reduction of feynman
  integrals.
\newblock {\em International Journal of Modern Physics C}, 27(03):1650027,
  2016.

\bibitem{pereira_2020}
Rajiv~G. Pereira.
\newblock Critical dynamics of the nonconserved strongly anisotropic
  permutation symmetric three-vector model.
\newblock {\em Phys. Rev. E}, 102:062150, Dec 2020.

\bibitem{Dutta_2011}
Sreedhar~B. Dutta and Su-Chan Park.
\newblock Critical dynamics of nonconserved $n$-vector models with anisotropic
  nonequilibrium perturbations.
\newblock {\em Phys. Rev. E}, 83:011117, Jan 2011.

\bibitem{hahn}
Thomas Hahn.
\newblock The feynarts visitor center: http://www.feynarts.de/.

\bibitem{FeynArtspaper}
J.~Küblbeck, M.~Böhm, and A.~Denner.
\newblock {\em Computer Physics Communications}, 60(2):165--180, 1990.

\bibitem{FeynArtspaper2}
Thomas Hahn.
\newblock Generating feynman diagrams and amplitudes with feynarts 3.
\newblock {\em Computer Physics Communications}, 140(3):418--431, 2001.

\bibitem{eck1995feynarts}
H~Eck.
\newblock {\em FeynArts 2.0—A generic Feynman diagram generator}.
\newblock PhD thesis, Universit{\"a}t W{\"u}rzburg, 1995.

\end{thebibliography}
\end{document}